\documentclass[12pt]{article}

\date{}
\begin{document}
\title{The matrices of argumentation frameworks}
\author{ {X\small{u} \large{Y}\small{uming}}
\thanks{\footnotesize  Corresponding author. E-mail: xuyuming@sdu.edu.cn }\\
{\small $School \ of \ Mathematics, Shandong \ University, Jinan,  China$}
 }
\maketitle
%\end{document}

\hspace{5mm} \begin{minipage}{12.5cm}

\begin{center}{\small \bf Abstract} \end{center}

{\small We introduce matrix and its block to the Dung's theory of argumentation frameworks. It is showed that each argumentation framework has a matrix representation, and the common extension-based semantics of argumentation framework can be characterized by blocks of matrix and their relations. In contrast with traditional method of directed graph, the matrix way has the advantage of computability. Therefore, it has an extensive perspective to bring the theory of matrix into the research of argumentation frameworks and related areas. }

\par {\small \it  \hspace{0cm} Keywords:}
{\small Argumentation framework; extension semantics; matrix; block}

\end{minipage}\\ \\

\hspace{-10mm} {\bf \large 1. Introduction}

\vspace{5mm}
In recent years, the area of argumentation begins to become increasingly central as a core study within Artificial Intelligence. A number of papers investigated and compared the properties of different semantics which have been proposed for argumentation frameworks (AFs, for short) as introduced by Dung [8, 4, 3, 9, 6]. In early time, many of the analysis of arguments are expressed in natural language. Later on, a tradition of using diagrams has been developed to explicate the relations between the components of the arguments. Now, argumentation frameworks are usually represented as directed graphs, which play a significant role in modeling and analyzing the extension-based semantics of AFs. For further notations and techniques of argumentation, we refer the reader to [8, 2, 15, 1].

Our aim is to introduce matrix as a new mathematic tool to the research of argumentation frameworks.
First, we assign a matrix of order $n$ for each argumentation framework with $n$ arguments. Each element of the matrix has only two possible values: one and zero, where one represents the attack relation and zero represents the non-attack relation between two arguments (they can be the same one). Under this circumstance, the matrix can be thought to be a representation of the argumentation framework. Secondly, we analysis the internal structure of the matrix corresponding to various extension-based semantics of the argumentation framework, and obtain the matrix approaches to determine the stable extension, admissible extension and complete extension, which can be easily realized on computer.

As will be seen in later, the matrix of an argumentation framework is not only visualized as the directed graph, but also has another significant advantage on the aspect of computation. We shall study various extension-based semantics of the argumentation framework by comparing and computing the matrix of the AF and its blocks. \\

\hspace{-10mm} {\bf \Large 2. Dung's theory of argumentation}

\vspace{5mm} Argumentation is a general approach to model defeasible reasoning and justification in Artificial Intelligence. So far, many theories of argumentation have been established. Among them, Dung's theory of argumentation framework is quite influence. In fact, it is abstract enough to manage without any assumption on the nature of arguments and the attack relation between arguments. Let us first recall some basic notion in Dung's theory of argumentation framework. We restrict them to finite argumentation frameworks.

An argumentation framework is a pair $F = (A, R)$, where $A$ is a finite set of arguments and $R \subset A \times A$ represents the attack-relation. For $S \subset A$, we say that

(1) $S$ is conflict-free in $(A, R)$ if there are no $a, b \in S$ such that $(a, b) \in R$;

(2) $a \in A$ is defeated by $S$ in $(A, R)$ if there is $b \in S$ such that $(b, a) \in R$;

(3) $a \in A$ is defended by $S$ in $(A, R)$ if for each $b \in A$ with $(b, a) \in R$,  we have $b$ is

\hspace{5mm} defeated by $S$ in $(A, R)$.

(4) $a \in A$ is acceptable with respect to $S$ if for each $b \in A$ with $(b, a) \in R$, there is some

\hspace{5mm} $c \in S$ such that $(c, b) \in R$.

The conflict-freeness, as observed by Baroni and Giacomin[1] in their study of evaluative criteria for extension-based semantics, is viewed as a minimal requirement to be satisfied within any computationally sensible notion of "collection of justified arguments". However, it is too weak a condition to be applied as a reasonable guarantor that a set of arguments is "collectively acceptable".

Semantics for argumentation frameworks can be given by a function $\sigma$ which assigns each AF $F = (A, R)$ a collection $\mathcal{S} \subset 2^{A}$ of extensions. Here, we mainly focus on the semantic $\sigma \in \{s, a, p, c, g, i, ss, e\}$ for stable, admissible, preferred, complete, grounded, ideal, semi-stable and eager extensions, respectively.

\hspace{-10mm} \textbf{Definition 1}[14] Let $F = (A, R)$ be an argumentation framework and $S \in A$.

(1) $S$ is a stable extension of $F$, $i.e.$, $S \in s(F)$, if $S$ is conflict-free in $F$ and each

\hspace{6mm} $a \in A \setminus S$ is defeated by $S$ in $F$.

(2) $S$ is an admissible extension of $F$,  $i.e.$, $S \in a(F)$, if $S$ is conflict-free in $F$ and each

\hspace{6mm} $a \in A \setminus S$ is defended by $S$ in $F$.

(3) $S$ is a preferred extension of $F$,  $i.e.$, $S \in p(F)$, if $S \in a(F)$ and for each $T \in a(F)$,

\hspace{6mm} we have $S \not\subset T$.

(4)  $S$ is a complete extension of $F$,  $i.e.$, $S \in c(F)$, if $S \in a(F)$ and for each $a \in A$

\hspace{6mm} defended by $S$ in $F$, we have $a \in S$.

(5)  $S$ is a grounded extension of $F$,  $i.e.$, $S \in g(F)$, if $S \in c(F)$ and for each $T \in c(F)$,

\hspace{6mm} we have $T \not\subset S$.

(6)  $S$ is an ideal extension of $F$,  $i.e.$, $S \in i(F)$, if $S \in a(F)$, $S \subset \cap \{T: T \in p(F)\}$ and

\hspace{6mm} for each $U \in a(F)$ such that $U \subset \cap \{T: T \in p(F)\}$, we have $S \not\subset U$.

(7)  $S$ is a semi-stable extension of $F$,  $i.e.$, $S \in ss(F)$, if $S \in a(F)$ and for each $T \in a(F)$,

\hspace{6mm} we have $R^{+}(S) \not\subset R^{+}(T)$, where $R^{+}(U) = \{U \cap \{b: (a, b) \in R, A \in U\}\}$.

(8)  $S$ is a eager extension of $F$,  $i.e.$, $S \in e(F)$, if $S \in c(F)$, $S \subset \cap \{T: T \in ss(F)\}$ and

\hspace{6mm} for each $U \in a(F)$ such that $U \subset \cap \{T: T \in ss(F)\}$, we have $S \not\subset T$.

\vspace{2mm} Note that, there are some elementary properties for any argumentation framework $F = (A, R)$ and semantic $\sigma$. If $\sigma \in \{a, p, c, g\}$, then we have $\sigma(F) \neq \emptyset$. And if $\sigma \in \{g, i, e\}$, then $\sigma(F)$ contains exactly one extension. Furthermore, the following relations hold for each argumentation framework $F = (A, R)$:

\vspace{2mm}\hspace{40mm}     $s(F) \subseteq p(F) \subseteq c(F) \subseteq a(F)$.

\vspace{2mm} Since every extension of an AF under the standard semantics (stable, preferred, complete and grounded extensions) introduced by Dung is an admissible set, the concept of admissible extensions plays an important role in the study of argumentation frameworks. \\

\hspace{-10mm} {\bf \Large 3.  The matrix of an argumentation framework}

\vspace{5mm} We know that the directed graph is a traditional tool in the research of argumentation frameworks, and has the feature of visualization [7, 10, 11]. It is widely used for modeling and analyzing argumentation frameworks. In this section, we shall introduce the matrix representation of argumentation frameworks. Except for the visualization, the matrix also has the advantage of computability in analyzing argumentation frameworks and computing various extension semantics.

An $m \times n$ matrix $A$ is a rectangular array of numbers, consisting of $m$ rows and $n$ columns, denoted by

\[
  A = \left(\begin{array}{cccccc}
  a_{1,1}&a_{1,2}&.&.&.&a_{1,n}\\
  a_{2,1}&a_{2,2}&.&.&.&a_{2,n}\\
  .&.&.&.&.&.\\
  a_{m,1}&a_{m,2}&.&.&.&a_{m,n}
  \end{array}\right).
\]
The $m \times n$ numbers $a_{1,1}, a_{1,2}, ..., a_{m,n}$ are the elements of the matrix $A$. We often called $a_{i,j}$ the $(i,j)$th element, and write $A = (a_{i, j})$ for short. It is important to remember that the first suffix of $a_{i,j}$ indicates the row and the second the column of $a_{i,j}$.

A column matrix is an $n \times 1$ matrix, and a row matrix is an $1 \times n$ matrix, denoted by

\[
  \left(\begin{array}{c}  x_{1}\\ x_{2}\\ .\\ .\\ .\\ x_{n}  \end{array}\right), \left(\begin{array}{cccccc}  x_{1}&x_{2}&.&.&.&x_{n}  \end{array}\right)
\]
respectively. Matrices of both these types can be regarded as vectors and referred to respectively as column vectors and row vectors.
Usually, the $i$th row of a matrix $A$ is denoted by $A_{i, *}$, and the $j$th column of $A$ is denoted by $A_{*, j}$.\\

\hspace{-10mm} \textbf{Definition 2} In an $n \times m$ matrix $A = (a_{i, j})$, we specify any $k ( \leq min\{n, m\} )$ different rows $i_{1},i_{2},...,i_{k}$ and the same number of different columns $i_{1},i_{2},...,i_{k}$. The elements appearing at the intersections of these rows and columns form a square matrix of order $k$. We call this matrix a principal block of order $k$ of the original matrix $A$; it is denoted by

\[
  M = \left(\begin{array}{cccccc}
  a_{i_{1}, i_{1}}&a_{i_{1}, i_{2}}&.&.&.&a_{i_{1}, i_{k}}\\
  a_{i_{2}, i_{1}}&a_{i_{2}, i_{2}}&.&.&.&a_{i_{2}, i_{k}}\\
  .&.&.&.&.&.\\
  a_{i_{k}, i_{1}}&a_{i_{k}, i_{2}}&.&.&.&a_{i_{k}, i_{k}}
  \end{array}\right),
\]
or $M = M_{i_{1},i_{2},...,i_{k}}^{i_{1},i_{2},...,i_{k}}$ for short.\\

\hspace{-10mm} \textbf{Definition 3} If in the original $n \times m$  matrix $A = (a_{i, j})$, we delete the rows and columns which make up the block $M = M^{i_{1},i_{2},...,i_{k}}_{i_{1},i_{2},...,i_{k}}$, then the remaining elements form an $(n-k) \times (m-k)$ matrix.
We call this matrix the complementary block of $M$, and is denoted by the symbol $\overline{M} = \overline{M_{i_{1},i_{2},...,i_{k}}^{i_{1},i_{2},...,i_{k}}}$. \\

\hspace{-10mm} \textbf{Definition 4} In an $n \times m$ matrix $A$, we specify any $k (\leq n)$ different rows $i_{1},i_{2},...,i_{k}$ and $h (\leq m)$ different columns $j_{1},j_{2},...,j_{h}$. The elements appearing at the intersections of these rows and columns form a $k \times h$ matrix. We call this matrix a $k \times h$ block of the original matrix $A$; it is denoted by

\[
  M = \left(\begin{array}{cccccc}
  a_{i_{1}, j_{1}}&a_{i_{1}, j_{2}}&.&.&.&a_{i_{1}, j_{h}}\\
  a_{i_{2}, j_{1}}&a_{i_{2}, j_{2}}&.&.&.&a_{i_{2}, j_{h}}\\
  .&.&.&.&.&.\\
  a_{i_{k}, j_{1}}&a_{i_{k}, j_{2}}&.&.&.&a_{i_{k}, j_{h}}
  \end{array}\right),
\]
or $M = M_{i_{1},i_{2},...,i_{k}}^{j_{1},j_{2},...,j_{h}}$ for short. \\

For the underlying set $A$ of an argumentation framework $F = (A, R)$, there is no ordering in nature. But, in many cases the ordering set can benefit us a lot. Contrasting with the form $A = \{a, b, ...\}$, it is more convenience to put $A = \{1, 2, ..., n\}$ while the cardinality of $A$ is large. In particular, we can map each argument to the corresponding row and column of a matrix. We will follow this arrangement in the below discussion.

\vspace{2mm}
\hspace{-10mm} \textbf{Definition 5} Let $F = (A, R)$ be an argumentation framework with $A = \{1, 2, ..., n\}$. The matrix of $F$, denoted by $M(F)$, is a Boolean matrix of order $n$, its element is determined by the following rules:

(1) $a_{i,j} = 1$ iff $(i, j) \in R$;

(2) $a_{i,j} = 0$ iff $(i, j) \notin R$.

\vspace{2mm}
\hspace{-10mm} \textbf{Example 6} Considering the argumentation framework $F = (A, R)$, where $A = \{1,2,3\}$ and $R = \{(1, 2), (2, 3), (3, 1)\}$. By the definition, we have the following matrix of $F$:

\[
  M(F) = \left(\begin{array}{ccc}
  0&1&0\\
  0&0&1\\
  1&0&0
  \end{array}\right)
\]

 \vspace{2mm}
\hspace{-10mm} \textbf{Example 7} Given an argumentation framework $F = (A, R)$, where $A = \{1,2,3,4\}$ and $R = \{(1, 2), (1, 3), (2, 1), (2, 3), (3, 4)\}$. The matrix of $F$ is as follows:

\[
  M(F) = \left(\begin{array}{cccc}
  0&1&1&0\\
  1&0&1&0\\
  0&0&0&1\\
  0&0&0&0
  \end{array}\right)
\]

In comparison with graph-theoretic way and mathematical logic way, the matrix of an argumentation framework has many excellent features. First, it possess a concise mathematical format. Secondly, it contains all information of the $AF$ by combining the arguments with attack relation in a specific manner in the matrix $M(F)$. Also, it can be deal with by program on computer. The most important is that we can import the knowledge of matrix to the research of argumentation frameworks. \\

\hspace{-10mm} {\bf \Large 4.  Determination of the conflict-free sets}

\vspace{5mm} As we know, there is no efficient method for us to decide a conflict set in an argumentation framework, even we can draw up the directed graph of the AF. After we introduce the matrix of the AF, the situation will be changed completely. By checking the matrix of the argumentation framework, we can easily find out all the conflict-free sets of the AF. Let us see an example, firstly.

\hspace{-10mm} \textbf{Example 8} Given an argumentation framework $F = (A, R)$, where $A = \{1,2,3,4,5\}$ and $R = \{(1, 2), (2, 3), (2, 5), (4, 3), (5, 4)\}$. Then, we can easily to show that the collection of conflict-free sets of $F$ is

     $\{ \emptyset, \{1\}, \{2\}, \{3\}, \{4\}, \{5\}£¬ \{1,3\}, \{1,4\}, \{1,5\}, \{2,4\}, \{3,5\}, \{1,3,5\} \}$,

\hspace{-10mm} by the routine method of directed graph.

On the other hand, we consider the matrix of $F = (A, R)$ and study its structure from the level of blocks. First, we write out the matrix of $F$:

\[
  M(F) = \left(\begin{array}{ccccc}
  0&1&0&0&0\\
  0&0&1&0&1\\
  0&0&0&0&0\\
  0&0&1&0&0\\
  0&0&0&1&0
  \end{array}\right).
\]

\hspace{-10mm} By observing the principal blocks of the above matrix, we find that there are five zero principal blocks of order 1
\[
  M^{1}_{1} = \left(\begin{array}{c}   0    \end{array} \right), M^{2}_{2} = \left(\begin{array}{c}  0  \end{array}  \right), M^{3}_{3} = \left(\begin{array}{c}   0  \end{array}  \right), M^{4}_{4} = \left(\begin{array}{c}   0  \end{array}  \right), M^{5}_{5} = \left(\begin{array}{c}   0  \end{array}  \right)
\]
corresponding to the conflict-free sets $\{1\}$, $\{2\}$, $\{3\}$, $\{4\}$, $\{5\}$, respectively.
There are five zero principal blocks of order 2
\[
  M^{1,3}_{1,3} = \left(\begin{array}{cc}   0&0\\    0&0    \end{array} \right), M^{1,4}_{1,4} = \left(\begin{array}{cc}   0&0\\    0&0    \end{array}  \right), M^{1,5}_{1,5} = \left(\begin{array}{cc}   0&0\\    0&0    \end{array}  \right), M^{2,4}_{2,4} = \left(\begin{array}{cc}   0&0\\    0&0    \end{array}  \right), M^{3,5}_{3,5} = \left(\begin{array}{cc}   0&0\\    0&0    \end{array}  \right)
\]
corresponding to the conflict-free sets $\{1,3\}$, $\{1,4\}$, $\{1,5\}$, $\{2,4\}$, $\{3,5\}$, respectively. Also, there is a zero principal block of order 3

\[
  M^{1,3,5}_{1,3,5} = \left(\begin{array}{ccc}   0&0&0\\    0&0&0\\   0&0&0    \end{array} \right)
\]
corresponding to the conflict-free sets $\{1, 3, 5\}$.

Note that, the above blocks are all principal blocks which are zero in the matrix $M(F)$, and there is a one to one correspond between the collection of all conflict-free sets of $F$ and the set of all zero principal blocks of $M(F)$. In fact, for any argumentation framework $F$ there exists such corresponding relation between the collection of all conflict-free sets of $F$ and the set of all zero principal blocks of $M(F)$.

 Since it is easy to find out the zero principal blocks in the matrix of an argumentation framework, we obtain a good way to decide the conflict-free sets of the $AF$ through its matrix. Certainly, this way can be carried out on the computer readily.

\vspace{2mm}
\hspace{-10mm} \textbf{Definition 9} Let $F = (A, R)$ be an argumentation framework with $A = \{1, 2, ..., n\}$, and $S = \{i_{1}, i_{2}, ..., i_{k}\} \subset A$. The principal block

\[
  M^{i_{1}, i_{2}, ..., i_{k}}_{i_{1}, i_{2}, ..., i_{k}} = \left(\begin{array}{cccccc}
  a_{i_{1}, i_{1}}&a_{i_{1}, i_{2}}&.&.&.&a_{i_{1}, i_{k}}\\
  a_{i_{2}, i_{1}}&a_{i_{2}, i_{2}}&.&.&.&a_{i_{2}, i_{k}}\\
  .&.&.&.&.&.\\
  a_{i_{k}, i_{1}}&a_{i_{k}, i_{2}}&.&.&.&a_{i_{k}, i_{k}}
  \end{array}\right)
\]
of order $k$ in the matrix $M(F)$ is called the $cf$-block of $S$, and denoted by $M^{cf}$.

\vspace{2mm}
\hspace{-10mm} \textbf{Theorem 10} Given an argumentation framework $F = (A, R)$ with $A = \{1, 2, ..., n\}$, then $S = \{i_{1}, i_{2}, ..., i_{k}\} \subset A$ is a conflict-free set in $F$ iff the $cf$-block $M^{i_{1}, i_{2}, ..., i_{k}}_{i_{1}, i_{2}, ..., i_{k}}$ of $S$ is zero.

\vspace{2mm}
\hspace{-10mm} \textbf{Proof} Assume that $M^{i_{1}, i_{2}, ..., i_{k}}_{i_{1}, i_{2}, ..., i_{k}} = 0$, then for arbitrary $1 \leq s, t \leq k$ we have $a_{i_{s}, i_{t}} = 0$, $i.e.$, $(i_{s}, i_{t}) \notin R$. Thus, $S = \{i_{1}, i_{2}, ..., i_{k}\}$ is a conflict-free set in $F$.

 Suppose $S = \{i_{1}, i_{2}, ..., i_{k}\} \subset A$ is a conflict-free set in $F$, then for arbitrary $1 \leq s, t \leq k$ we have that $(i_{s}, i_{t}) \notin R$, $i.e.$, $a_{i_{s}, i_{t}} = 0$. Therefore, we have $M^{i_{1}, i_{2}, ..., i_{k}}_{i_{1}, i_{2}, ..., i_{k}} = 0$.\\

\hspace{-10mm} {\bf \Large 5.  Determination of the stable extensions}

\vspace{5mm}
\hspace{-10mm} \textbf{Example 11} We continuous to study the argumentation framework $F = (A, R)$, where $A = \{1,2,3,4,5\}$ and $R = \{(1, 2), (2, 3), (2, 5), (4, 3), (5, 4)\}$. Since the stable extension is firstly a conflict-free set, we can look for the stable extension from the collection

$\{\emptyset, \{1\}, \{2\}, \{3\}, \{4\}, \{5\}£¬\{1,3\}, \{1,4\}, \{1,5\}, \{2,4\}, \{3,5\}, \{1,3,5\} \}$

\hspace{-10mm} of conflict-free sets. In fact, the set $S = \{1,3,5\}$ is the only stable extension in $F$ by a simple discussion.

Again, we turn our attention to the matrix of the $F = (A, R)$:

\[
  M(F) = \left(\begin{array}{ccccc}    0&1&0&0&0\\      0&0&1&0&1\\     0&0&0&0&0\\     0&0&1&0&0\\     0&0&0&1&0     \end{array}\right).
\]

\vspace{2mm}\hspace{-10mm} Since $S = \{1,3,5\}$ is a stable extension of $F$, the arguments $2$ and $4$ are defeated by $\{1, 3, 5\}$. This fact is reflected in the matrix $M(F)$ of $F$ as follows.

In the column vector $F_{*,2}$ (column 2), $a_{1, 2} = 1$ means that $(1, 2) \in R$, and thus the argument $1$ attacks the argument $2$. In the column vector $F_{*,4}$ (column 4), $a_{5, 4} = 1$ means that $(5, 4) \in R$, and thus the argument $5$ attacks the argument $4$.

From the behavior of the elements $a_{1, 2} = 1$ and $a_{5, 4} = 1$ in the matrix $M(F)$, we can extract a matrix approach to decide that the conflict-free set $S = \{1, 3, 5\}$ is a stable extension: Corresponding to the arguments $2, 4 \in A \setminus S$, we firstly pick out the column vectors $F_{*,2}$ and $F_{*,4}$ in the matrix $M(F)$, then check the elements $a_{1, 2}, a_{3, 2}, a_{5, 2}$ of $F_{*, 2}$, and the elements $a_{1, 4}, a_{3, 4}, a_{5, 4}$ of $F_{*, 4}$. If there is one element of $\{a_{1, 2}, a_{3, 2}, a_{5, 2}\}$ which is non-zero, then the argument $2$ is defeated by $S$. Similar result is hold for the argument $4$. This process leads to a block of the matrix $M(F)$ at the intersection of columns $2, 4$ and rows $1, 3, 5$.

To sum up, we can decide that the conflict set $S = \{1, 3, 5\}$ is a stable extension by the fact that the two column vectors of the above block of the matrix $M(F)$ are all non-zero.
Further analysis indicates that the converse is also true. This motivation makes us to give the following definition.

\vspace{2mm}
\hspace{-10mm} \textbf{Definition 12} Let $F = (A, R)$ be an argumentation framework with $A = \{1, 2, ..., n\}$, and $S = \{i_{1}, i_{2}, ..., i_{k}\} \subset A$ is a stable extension of $F$. The $k \times h$ block

\[
  M^{i_{1}, i_{2}, ..., i_{k}}_{j_{1}, j_{2}, ..., j_{h}} = \left(\begin{array}{cccccc}
  a_{i_{1}, j_{1}}&a_{i_{1}, j_{2}}&.&.&.&a_{i_{1}, j_{h}}\\
  a_{i_{2}, j_{1}}&a_{i_{2}, j_{2}}&.&.&.&a_{i_{2}, j_{h}}\\
  .&.&.&.&.&.\\
  a_{i_{k}, j_{1}}&a_{i_{k}, j_{2}}&.&.&.&a_{i_{k}, j_{h}}
  \end{array}\right)
\]
in the matrix $M(F)$ is called the $s$-block of $S$ and denoted by $M^{s}$, where $\{j_{1}, j_{2}, ..., j_{h}\} = A \setminus S$.

In other words, the elements appearing at the intersections of rows $i_{1}, i_{2}, ..., i_{k}$ and columns $j_{1}, j_{2}, ..., j_{h}$ in the matrix $M(F)$ form the $s$-block $M^{i_{1}, i_{2}, ..., i_{k}}_{j_{1}, j_{2}, ..., j_{h}}$ of $S$.

\vspace{2mm}
\hspace{-10mm} \textbf{Theorem 13} Given an argumentation framework $F = (A, R)$ with $A = \{1, 2, ..., n\}$, then $S = \{i_{1}, i_{2}, ..., i_{k}\}  \subset A$ is a stable extension in $F$ iff the following conditions hold:

(1) The $cf$-block $M^{i_{1}, i_{2}, ..., i_{k}}_{i_{1}, i_{2}, ..., i_{k}}$ of $S$ is zero,

(2) Every column vector of the $s$-block $M^{i_{1}, i_{2}, ..., i_{k}}_{j_{1}, j_{2}, ..., j_{h}}$ of $S$ is non-zero, where $A \setminus S$

\hspace{6mm}  $ = \{j_{1}, j_{2}, ..., j_{h}\}$.

\vspace{2mm}
\hspace{-10mm} \textbf{Proof} Let $S$ be a conflict-free set and $A \setminus S = \{j_{1}, j_{2}, ..., j_{h}\}$, then we need only to prove that every element of $A \setminus S (1 \leq t \leq h)$ is defeated by $S$ in $F$ iff all column vectors of the $s$-block $M^{i_{1}, i_{2}, ..., i_{k}}_{j_{1}, j_{2}, ..., j_{h}}$ of $S$ are non-zero.

  Assume that every element of $A \setminus S (1 \leq t \leq h)$ is defeated by $S$ in $F$. Take any column vector $A_{*, j_{t}} (1 \leq t \leq h)$ of the $s$-block $M^{i_{1}, i_{2}, ..., i_{k}}_{j_{1}, j_{2}, ..., j_{h}}$ of $S$, then we have $j_{t} \in A \setminus S$. By the assumption, there is some element $i_{r} \in S (1 \leq r \leq k)$ such that the argument $i_{r}$ attacks the argument $j_{t}$, $i.e.$, $(i_{r}, j_{t}) \in R$. It follows that $a_{i_{r}, j_{t}} = 1$ in the matrix $M(F)$ and the $s$-block $M^{i_{1}, i_{2}, ..., i_{k}}_{j_{1}, j_{2}, ..., j_{h}}$ of $S$, and thus the column vector $A_{*, j_{t}}$ is non-zero.

Conversely, suppose that all column vectors of the $s$-block $M^{i_{1}, i_{2}, ..., i_{k}}_{j_{1}, j_{2}, ..., j_{h}} = M^{s}$ of $S$ are non-zero. Take any element $j_{t} \in A \setminus S (1 \leq t \leq h)$, then $M^{s}_{*, j_{t}}$ is a column vector of the $s$-block $M^{i_{1}, i_{2}, ..., i_{k}}_{j_{1}, j_{2}, ..., j_{h}} = M^{s}$ of $S$. By the hypothesis, we know that $A_{*, j_{t}}$ is non-zero. Therefore, there is some $i_{r} \in S (1 \leq r \leq k)$ such that $a_{i_{r}, j_{t}} = 1$, $i.e.$, $(i_{r}, j_{t}) \in R$. This means that the argument $i_{r}$ attacks the argument $j_{t}$ of $S$ in $F$, and thus we claim that $j_{t}$ is defeated by $S$ in $F$. \\

\hspace{-10mm} {\bf \Large 6.  Determination of the admissible extensions}

\vspace{5mm}
\hspace{-10mm} \textbf{Example 14} Let us return to the argumentation framework $F = (A, R)$, where $A = \{1,2,3,4,5\}$ and $R = \{(1, 2), (2, 3), (2, 5), (4, 3), (5, 4)\}$. Since an admissible extension is necessarily a conflict-free set, we can look for the admissible extension from the collection

$\{\emptyset, \{1\}, \{2\}, \{3\}, \{4\}, \{5\}£¬\{1,3\}, \{1,4\}, \{1,5\}, \{2,4\}, \{3,5\}, \{1,3,5\} \}$

\hspace{-10mm} of conflict-free sets. By definition, it is easy to check that $\{1\}$, $\{1,5\}$ and $\{1,3,5\}$ are all the admissible extensions in $F$.

Since $\{1,3,5\}$ is also a stable extension and $\{1\}$ is not typical enough as an admissible extension in $F$, we will mainly concentrate on the admissible extension $S = \{1,5\}$ which is not a stable extension in $F$.

First, we write out the matrix of argumentation framework $F = (A, R)$:

\[
  M(F) = \left(\begin{array}{ccccc}    0&1&0&0&0\\      0&0&1&0&1\\     0&0&0&0&0\\     0&0&1&0&0\\     0&0&0&1&0     \end{array}\right).
\]

Secondly, we study the structure of the matrix $M(F)$ of $F$ to find out the internal properties which can reflect the fact that $S = \{1,5\}$ is an admissible extension.

In the column vector $M(F)_{*, 5}$ of the matrix $M(F)$, $a_{2,5} = 1$ means that $(2, 5) \in R$, $i.e.$, the argument $2$ attacks the argument $5$. Under this circumstance, the element $a_{1, 2} = 1$ in the row vector $M(F)_{*, 2}$ of the matrix $M(F)$ implies that $(1, 2) \in R$, $i.e.$,  the argument $1$ attacks the argument $2$. This illustrates that the argument $5$ is defended by $\{1,5\}$ in $F$. In the column vector $M(F)_{*, 1}$ of the matrix $M(F)$, we have $a_{i, 1} = 0$ for each $1 \leq i \leq 5$. It follows that the argument $1$ is defended by $\{1,5\}$ in $F$.

In the above analysis, the behavior of $a_{2, 5} = 1$ and $a_{1, 2} = 1$ in the matrix $M(F)$ is intrinsic for the fact that the argument $5$ is defended by $\{1,5\}$ in $F$. This inspires us a general idea to decide the conflict-free set $S = \{1, 5\}$ to be admissible through the structure of the matrix $M(F)$ of $F$.

\vspace{2mm} (1) In order to decide whether the arguments of $\{1, 5\} = S$ are defended by $S$, we should firstly find the attackers of the argument $1$ and $5$.  So, we must pick out the column vectors $M(F)_{*, 1}$ and $M(F)_{*, 5}$ of the matrix $M(F)$ corresponding to the arguments $1$ and $5$ respectively. Since the set $S$ is conflict-free, there is no attack relation between $1$ and $5$, $i.e.$, $a_{1, 1} = 0, a_{5, 1} = 0, a_{1, 5} = 0, a_{5, 5} = 0$. Therefore, we only need to check the elements $a_{2, 1}, a_{3, 1}, a_{4, 1}$ of the column vector $M(F)_{*, 1}$, and the elements  $a_{2, 5},a_{3, 5}, a_{4, 5}$ of the column vector $M(F)_{*, 5}$. Each non-zero element of the set $\{a_{2, 1}, a_{3, 1}, a_{4, 1}\}$ tells us an attacker of the argument $1$, and each non-zero element of the set $\{a_{2, 5}, a_{3, 5}, a_{4, 5}\}$ tells us an attacker of the argument $5$. This leads to a block of the matrix $M(F)$ at the intersection of column $1, 5$ and row $2, 3, 4$, which is exactly the $s$-block of $S$.

(2) After having determined the attackers $(\in \{2, 3, 4\})$ of the argument $1$ and $5$, we should secondly to check whether these attackers are defeated by $S = \{1, 5\}$. For example, $a_{2, 5} = 1$ means that the argument $2$ is an attacker of the argument $5$. So, we should check the element $a_{1, 2}$ and $a_{5, 2}$ to see whether the attacker $2$ of the argument $5$ is defeated by $\{1, 5\}$. Similar situation holds for any other attackers of the argument $1$ and $5$. Namely, we need also to check the elements $a_{1, 3}, a_{5, 3}$ ( if the argument $3$ is an attacker of the argument $1$ or $5$ ) and elements $a_{1, 4}, a_{5, 4}$ ( if the argument $4$ is an attacker of the argument $1$ or $5$). This process leads to a block of the matrix $M(F)$ at the intersection of columns $2, 3, 4$ and rows $1, 5$.

In summary, we need to check two blocks (related to $S = \{1, 5\}$) of the matrix $M(F)$ in order to decide that the conflict-free set $S = \{1, 5\}$ is an admissible extension. This motivate us to give the following definition.

\vspace{2mm}
\hspace{-10mm} \textbf{Definition 15} Let $F = (A, R)$ be an argumentation framework with $A = \{1, 2, ..., n\}$, and $S = \{i_{1}, i_{2}, ..., i_{k}\} \subset A$ is an admissible extension of $F$. The $h \times k$ block

\[
  M^{j_{1}, j_{2}, ..., j_{h}}_{i_{1}, i_{2}, ..., i_{k}} = \left(\begin{array}{cccccc}
  a_{j_{1}, i_{1}}&a_{j_{1}, i_{2}}&.&.&.&a_{j_{1}, i_{k}}\\
  a_{j_{2}, i_{1}}&a_{j_{2}, i_{2}}&.&.&.&a_{j_{2}, i_{k}}\\
  .&.&.&.&.&.\\
  a_{j_{h}, i_{1}}&a_{j_{h}, i_{2}}&.&.&.&a_{j_{h}, i_{k}}
  \end{array}\right)
\]
of the matrix $M(F)$ is called the $a$-block of $S$ and denoted by $M^{a}$, where $\{j_{1}, j_{2}, ..., j_{h}\} = A \setminus S$.

In other words, the elements appearing at the intersection of rows $j_{1}, j_{2}, ..., j_{h}$ and columns $i_{1}, i_{2}, ..., i_{k}$ in the matrix $M(F)$ form the $a$-block $M^{j_{1}, j_{2}, ..., j_{h}}_{i_{1}, i_{2}, ..., i_{k}}$ of $S$.

Note that, there is a natural relation between the $a$-block $M^{j_{1}, j_{2}, ..., j_{h}}_{i_{1}, i_{2}, ..., i_{k}}$ and the $s$-block $M^{i_{1}, i_{2}, ..., i_{k}}_{j_{1}, j_{2}, ..., j_{h}}$ in matrix theory. Namely, the $a$-block $M^{j_{1}, j_{2}, ..., j_{h}}_{i_{1}, i_{2}, ..., i_{k}}$ of $S$ is precisely the complementary block of the $s$-block $M^{i_{1}, i_{2}, ..., i_{k}}_{j_{1}, j_{2}, ..., j_{h}}$  of $S$ in the matrix $M(F)$.

For convenience, in this section we may assume that the sequences $i_{1}, i_{2}, ..., i_{k}$ and $j_{1}, j_{2}, ..., j_{h}$ are all increasing.

\vspace{2mm}
\hspace{-10mm} \textbf{Theorem 16} Given an argumentation framework $F = (A, R)$ with $A = \{1, 2, ..., n\}$, then $S = \{i_{1}, i_{2}, ..., i_{k}\} \subset A$ is an admissible extension in $F$ iff the following conditions hold:

(1) The $cf$-block $M^{i_{1}, i_{2}, ..., i_{k}}_{i_{1}, i_{2}, ..., i_{k}}$ of $S$ is zero,

 (2) The column vector of $s$-block $M^{i_{1}, i_{2}, ..., i_{k}}_{j_{1}, j_{2}, ..., j_{h}}$ of $S$ corresponding to the non-zero row

\hspace{6mm} vector of the $a$-block $M^{j_{1}, j_{2}, ..., j_{h}}_{i_{1}, i_{2}, ..., i_{k}}$ of $S$ is non-zero, where $A \setminus S = \{j_{1}, j_{2}, ..., j_{h}\}$.

\vspace{2mm}
\hspace{-10mm} \textbf{Proof} Let $S$ be a conflict-free set and $A \setminus S = \{j_{1}, j_{2}, ..., j_{h}\}$. We need only to prove that every $i_{r} \in S (1 \leq r \leq k)$ is defended by $S$ in $F$ iff the column vector of $s$-block $M^{i_{1}, i_{2}, ..., i_{k}}_{j_{1}, j_{2}, ..., j_{h}}$ of $S$ corresponding to the non-zero row vector of the $a$-block $M^{j_{1}, j_{2}, ..., j_{h}}_{i_{1}, i_{2}, ..., i_{k}}$ of $S$ is non-zero

Assume that every $i_{r} \in S (1 \leq r \leq k)$ is defended by $S$ in $F$. If the row vector $M^{a}_{t, *} (1 \leq t \leq h)$ of the $a$-block $M^{j_{1}, j_{2}, ..., j_{h}}_{i_{1}, i_{2}, ..., i_{k}} = M^{a}$ of $S$ is non-zero, then there is some $i_{r} (1 \leq r \leq k)$ such that $a_{j_{t}, i_{r}} =1$. Note that $a_{j_{t}, i_{r}}$ is at the intersection of row $t$ and column $r$ of the $a$-block $M^{a}$ of $S$, and at the intersection of row $j_{t}$ and column $i_{r}$ of the matrix $M(F)$. This implies that $(j_{t}, i_{r}) \in R$, $i.e.$, the argument $j_{t}$ attacks the argument $i_{r}$. By the assumption, there is some $i_{q} \in S (1 \leq q \leq k)$ such that the argument $i_{q}$ attacks the argument $j_{t}$, $i.e.$, $(i_{q}, j_{t}) \in R$. It follows that $a_{i_{q}, j_{t}} = 1$ in the matrix $M(F)$. But, $a_{i_{q}, j_{t}}$ is also an element of the $s$-block $M^{s}$, which is at the intersection of row $q$ and column $t$ of $M^{s}$. Namely, $a_{i_{q}, j_{t}}$ is an element of the column vector $M^{s}_{*, t}$ of $M_{s}$. Therefore, we conclude that the column vector $M^{s}_{*, t}$ of $s$-block $M^{i_{1}, i_{2}, ..., i_{k}}_{j_{1}, j_{2}, ..., j_{h}} = M^{s}$ of $S$ is non-zero.

Conversely, suppose that the column vector of $s$-block $M^{i_{1}, i_{2}, ..., i_{k}}_{j_{1}, j_{2}, ..., j_{h}}$ of $S$ corresponding to the non-zero row vector of the $a$-block $M^{j_{1}, j_{2}, ..., j_{h}}_{i_{1}, i_{2}, ..., i_{k}}$ of $S$ is non-zero. For any fixed $i_{r} \in S (1 \leq r \leq k)$, if there is no $j_{t} \in A \setminus S (1 \leq t \leq h)$ such that the argument $j_{t}$ attacks the argument $i_{r}$, then by the fact that $S$ is a conflict-free set we claim that there is no $i \in A$ such that the argument $i$ attacks the argument $i_{r}$. It follows that argument $i_{r} \in S$ is defended by $S$ in $F$.

Otherwise, there is some $j_{t} \in A \setminus S (1 \leq t \leq h)$ such that the argument $j_{t}$ attacks the argument $i_{r}$. It follows that $(j_{t}, i_{r}) \in R$, $i.e.$, $a_{j_{t}, i_{r}} = 1$. Since the element $a_{j_{t}, i_{r}}$ is at the intersection of row $t$ and column $r$ of the $a$-block $M^{j_{1}, j_{2}, ..., j_{h}}_{i_{1}, i_{2}, ..., i_{k}} = M^{a}$ of $S$, the row vector $M^{a}_{t, *}$ of the $a$-block $M^{a}$ of $S$ is non-zero. By the assumption, we conclude that the corresponding column vector $M^{s}_{*, t}$ of the $s$-block $M^{i_{1}, i_{2}, ..., i_{k}}_{j_{1}, j_{2}, ..., j_{h}} = M^{s}$ of $S$ is non-zero. Therefore, there is some $i_{q} \in S (1 \leq q \leq k)$ such that $a_{i_{q}, j_{t}} = 1$. Note that, the element $a_{i_{q}, j_{t}}$ is at the intersection of row $q$ and column $t$ of the $s$-block $M^{i_{1}, i_{2}, ..., i_{k}}_{j_{1}, j_{2}, ..., j_{h}}$ and at the intersection of row $i_{q}$ and column $j_{t}$ of the matrix $M(F)$. Consequently, we have that $(i_{q}, j_{t}) \in R$, $i.e.$, the argument $i_{q} \in S$ attacks the argument $j_{t}$. Now, we have proved that the argument $i_{r} \in S$ is also defended by $S$ in $F$.

Remark: The fact that any stable extension must be admissible is clearly expressed by the properties of $s$-blocks in the matrix. In other words, the condition every column vector of the $s$-block $M^{i_{1}, i_{2}, ..., i_{k}}_{j_{1}, j_{2}, ..., j_{h}}$ of $S$ are non-zero is stronger than that the column vector of the $s$-block $M^{i_{1}, i_{2}, ..., i_{k}}_{j_{1}, j_{2}, ..., j_{h}}$ of $S$ corresponding to the non-zero row vector of the $a$-block $M^{j_{1}, j_{2}, ..., j_{h}}_{i_{1}, i_{2}, ..., i_{k}}$ of $S$ is non-zero. \\

\hspace{-10mm} {\bf \Large 7.  Determination of the complete extensions}

\vspace{5mm}
\hspace{-10mm} \textbf{Example 17} Consider the argumentation framework $F = (A, R)$, where $A = \{1,2,3,4,5\}$ and $R = \{(1, 2), (2, 3), \{2,4\}, (2, 5), (4, 3), (5, 4)\}$. Since the admissible extension is necessarily a conflict-free set, we can find out the admissible extension from the collection of conflict-free sets

$\{\emptyset, \{1\}, \{2\}, \{3\}, \{4\}, \{5\}£¬\{1,3\}, \{1,4\}, \{1,5\}, \{3,5\}, \{1,3,5\} \}$.

\hspace{-10mm}  By the directed graph of $F$, it is easy to check that $\{1,5\}$ and $\{1,3,5\}$ are all the admissible extensions in $F$. Furthermore, one can verify that $S_{1} = \{1,3,5\}$ is the only complete extension in $F$, while $S_{2} = \{1,5\}$ is not.

Next, we will analysis the different expressions in the matrix $M(F)$ of $F$ between $\{1,3,5\}$ (as a complete extension but not an admissible extension) and $\{1,5\}$ (as an admissible extension). By comparing them, we extract the matrix method to decide that an admissible extension is complete.

Let us firstly write out the matrix of the argumentation framework $F$:

\[
  M(F) = \left(\begin{array}{ccccc}    0&1&0&0&0\\      0&0&1&1&1\\     0&0&0&0&0\\     0&0&1&0&0\\     0&0&0&1&0     \end{array}\right).
\]

In the column vector $M(F)_{*, 2}$ of the matrix $M(F)$, $a_{1,2} = 1$ means that $(1, 2) \in R$, $i.e.$, the argument $1$ attacks the argument $2$. Since $S_{1} = \{1, 3, 5\}$ is a conflict-free set, there is no element of $S_{1}$ which attacks the argument $1$. It follows that the arguments $2$ is not defended by $S_{1}$ in $F$. In the column vector $M(F)_{*, 4}$ of the matrix $M(F)$, $a_{5, 4} = 1$ means that $(5, 4) \in R$, $i.e.$, the argument $5$ attacks the argument $4$. Also because that $S_{1} = \{1, 3, 5\}$ is a conflict-free set, there is no element of $S_{1}$ which attacks the argument $5$. Thus, we have that the arguments $4$ is not defended by $S_{1}$ in $F$. These are exactly the reasons for the admissible extension $S_{1} = \{1,3,5\}$ to be a complete extension.

Next, we will mainly focus our attention on the argument $3$ with respect to $S_{2} = \{1,5\}$.

In the column vector $M(F)_{*, 3}$ of the matrix $M(F)$, $a_{2,3} = 1$ means that $(2, 3) \in R$, and $a_{4,3} = 1$ means that  $(4, 3) \in R$. Therefore, both arguments $2$ and $4$ attack the argument $3$. On the other hand, in the column vector $M(F)_{*, 2}$ of the matrix $M(F)$, $a_{1, 2} = 1$ means that $(1, 2) \in R$, $i.e.$, the argument $1$ attacks the argument $2$. In the column vector $M(F)_{*, 4}$ of the matrix $M(F)$, $a_{5, 4} = 1$ means that $(5, 4) \in R$, $i.e.$, the argument $5$ attacks the argument $4$. Consequently, we have that the argument $3$ is defended by $S_{2} = \{1,5\}$ in $F$. It is precisely that the argument $3$ is not included in $S_{2}$ which leads to the fact that $S_{2} = \{1,5\}$ is not a complete extension.

From the above analysis, we find a simple fact: In an argumentation framework $F = (A, R)$ with $A = \{1, 2, ..., n\}$, an admissible extension $S = \{i_{1}, i_{2}, ..., i_{k}\}$ is complete iff each argument of $A \setminus S = \{j_{1}, j_{2}, ..., j_{h}\}$ is not defended by $S$ in $F$. And, we can summarize the process to decide an admissible extension $S$ to be complete by the blocks of matrix $M(F)$ of $F$ as follows:

(1) First, we pick out the column vectors $M(F)_{\ast, j_{1}}, M(F)_{\ast, j_{2}}, ..., M(F)_{\ast, j_{h}}$ of the matrix $M(F)$ corresponding to the arguments of $A \setminus S =  \{j_{1}, j_{2}, ..., j_{h}\}$. For each argument $j_{t} \in A \setminus S (1 \leq t \leq h)$, we check the elements $a_{1, j_{t}}, a_{2, j_{t}}, ..., a_{n, j_{t}}$ in the column vector $M(F)_{\ast, j_{t}}$ of the matrix $M(F)$ to find all the attackers of the argument $j_{t}$.

(2) For each argument $j_{t} (1 \leq t \leq h)$, we consider two cases with respect to its attackers.

\hspace{-10mm} (a) There is some $j_{p} \in A \setminus S (1 \leq p \leq h)$ such that $a_{j_{p}, j_{t}} = 1$ in the column vector $M(F)_{\ast, j_{t}}$ of the matrix $M(F)$, $i.e.$, $(j_{p}, j_{t}) \in R$, then the argument $j_{p}$ attacks the argument $j_{t}$ in $F$. In order that the argument $j_{t}$ is not defended by $S$, any argument $i_{r} \in S (1 \leq r \leq k)$ should not attack the argument $j_{p}$. Thus, we have $(i_{r}, j_{p}) \notin R$, $i.e.$, $a_{i_{r}, j_{p}} = 0$ for all $1 \leq r \leq k$.

\hspace{-10mm} (b) There is no $j_{p} \in A \setminus S (1 \leq p \leq h)$ such that $a_{j_{p}, j_{t}} = 1$ in the column vector $M(F)_{\ast, j_{t}}$ of the matrix $M(F)$, then there must be some $i_{r} \in S (1 \leq r \leq k)$ such that $a_{i_{r}, j_{t}} = 1$ in the column vector $M(F)_{\ast, j_{t}}$. Otherwise, there is no $i \in A$ such that $a_{i, j_{t}} = 1$, $i.e.$, there is no $i \in A$ such that $(i, j_{t}) \in R$. It follows that there is no argument $i \in A$ which attacks the argument $j_{t}$ in $F$.  This implies that the argument $j_{t}$ is defended by $S$  in $F$, and thus $S$ is not a complete extension.

In case $(a)$, the elements $"a_{j_{p}, j_{t}}" (1 \leq p \leq h, 1 \leq t \leq h)$ form a block of the matrix $M(F)$ at the intersection of row $j_{1}, j_{2}, ..., j_{h}$ and the same number of columns. The elements $"a_{i_{r}, j_{t}}" (1 \leq r \leq k, 1 \leq t \leq h)$ form anther block of the matrix $M(F)$ at the intersection of row $i_{1}, i_{2}, ..., i_{k}$ and the column $j_{1}, j_{2}, ..., j_{h}$, which is exactly the $s$-block of $S$. In case $(b)$, one can find that the elements considered form the same  blocks as in case $(a)$.
This motivation makes us to give the following definition.

\vspace{2mm}
\hspace{-10mm} \textbf{Definition 18} Let $F = (A, R)$ be an argumentation framework with $A = \{1, 2, ..., n\}$, and $S = \{i_{1}, i_{2}, ..., i_{k}\} \subset A$ is a complete extension of $F$. The block

\[
  M^{j_{1}, j_{2}, ..., j_{h}}_{j_{1}, j_{2}, ..., j_{h}} = \left(\begin{array}{cccccc}
  a_{j_{1}, i_{1}}&a_{j_{1}, i_{2}}&.&.&.&a_{j_{1}, i_{k}}\\
  a_{j_{2}, i_{1}}&a_{j_{2}, i_{2}}&.&.&.&a_{j_{2}, i_{k}}\\
  .&.&.&.&.&.\\
  a_{j_{h}, i_{1}}&a_{j_{h}, i_{2}}&.&.&.&a_{j_{h}, i_{k}}
  \end{array}\right)
\]
of order $h$ in the matrix of $M(F)$ is called the $c$-block of $S$ and denoted by $M^{c}$, where $\{j_{1}, j_{2}, ..., j_{h}\} = A \setminus S$.

In other words, the elements appearing at the intersection of rows $j_{1}, j_{2}, ..., j_{h}$ and the same number of columns in the matrix $M(F)$ form the $c$-block $M^{j_{1}, j_{2}, ..., j_{h}}_{j_{1}, j_{2}, ..., j_{h}}$ of $S$.

Note that, the $c$-block $M^{c} = M^{j_{1}, j_{2}, ..., j_{h}}_{j_{1}, j_{2}, ..., j_{h}}$ of $S$ is exactly the complementary block of the $s$-block $M^{s} = M^{i_{1}, i_{2}, ..., i_{k}}_{i_{1}, i_{2}, ..., i_{k}}$ of $S$, in the matrix $M(F)$ of $F$.

Now, the fact that $S_{1} = \{1, 3, 5\}$ is a complete extension in the above example can be verified by the following conditions:

(1) The column vector of $s$-block $M^{1, 3, 5}_{2, 4}$ of $S_{1}$ corresponding to the non-zero row vector

\hspace{6mm} of $c$-block $M^{2, 4}_{2, 4}$ of $S_{1}$ is zero;

(2) The column vector of $s$-block $M^{1, 3, 5}_{2, 4}$ of $S_{1}$ corresponding to the zero column vector

\hspace{6mm} of $c$-block $M^{2, 4}_{2, 4}$ of $S_{1}$ is non-zero.

For convenience, in this section we also assume that the sequences $i_{1}, i_{2}, ..., i_{k}$ and $j_{1}, j_{2}, ..., j_{h}$ are all increasing.

\vspace{2mm}
\hspace{-10mm} \textbf{Lemma 19} Let $F = (A, R)$ be an argumentation framework with $A = \{1, 2, ..., n\}$, then $S = \{i_{1}, i_{2}, ..., i_{k}\} \subset A$ is a complete extension of $F$ iff $S$ is an admissible extension and each argument $j_{t} \in S (1 \leq t \leq h)$ is not defended by $S$ in $F$.

\vspace{2mm}
\hspace{-10mm} \textbf{Theorem 20} Given an argumentation framework $F = (A, R)$ with $A = \{1, 2, ..., n\}$, then the admissible extension $S = \{i_{1}, i_{2}, ..., i_{k}\} \subset A$ is a complete extension in $F$ iff the following conditions hold:

(1) the column vector of $s$-block $M^{i_{1}, i_{2}, ..., i_{k}}_{j_{1}, j_{2}, ..., j_{h}}$ of $S$ corresponding to the non-zero row vector

\hspace{6mm} of the $c$-block $M^{j_{1}, j_{2}, ..., j_{h}}_{j_{1}, j_{2}, ..., j_{h}}$ of $S$ is zero,

(2) the column vector of $s$-block $M^{i_{1}, i_{2}, ..., i_{k}}_{j_{1}, j_{2}, ..., j_{h}}$ of $S$ corresponding to the zero column vector

\hspace{6mm} of the $c$-block $M^{j_{1}, j_{2}, ..., j_{h}}_{j_{1}, j_{2}, ..., j_{h}}$ of $S$ is non-zero,

\hspace{-10mm} where $A \setminus S = \{j_{1}, j_{2}, ..., j_{h}\}$.

\vspace{2mm}
\hspace{-10mm} \textbf{Proof} Let $S$ be an admissible extension and $A \setminus S = \{j_{1}, j_{2}, ..., j_{h}\}$, we need only to prove that every $j_{t} \in S (1 \leq t \leq h)$ is not defended by $S$ in $F$ iff the condition $(1)$ and $(2)$ are hold.

Assume that every $j_{t} \in A \setminus S (1 \leq t \leq h)$ is not defended by $S$ in $F$. If the row vector $M^{^{c}}_{r, *} (1 \leq r \leq h)$ of the $c$-block $M^{j_{1}, j_{2}, ..., j_{h}}_{j_{1}, j_{2}, ..., j_{h}}$ of $S$ is non-zero, then there is some $1 \leq t \leq h$ such that $a_{j_{r}, j_{t}} = 1$, $i.e.$, $(j_{r}, j_{t}) \in R$. It follows that the argument $a_{j_{r}}$ attacks the argument $a_{j_{t}}$. By the assumption, there is no argument in $S$ which attacks the argument $a_{j_{r}}$. Therefore, for each $i_{q} \in S (1 \leq q \leq k)$ we have $(i_{q}, j_{r}) \notin R$, $i.e.$, $a_{i_{q}, j_{r}} = 0$. This means that the column vector $M^{^{s}}_{*, r}$ of the $s$-block $M^{i_{1}, i_{2}, ..., i_{k}}_{j_{1}, j_{2}, ..., j_{h}}$ of $S$ is zero.

If the column vector $M^{^{c}}_{*, t} (1 \leq t \leq h)$  of the $c$-block $M^{j_{1}, j_{2}, ..., j_{h}}_{j_{1}, j_{2}, ..., j_{h}}$ of $S$ is zero, then for each $1 \leq p \leq h$ we have that $a_{j_{p}, j_{t}} = 0$, $i.e.$, $(j_{p}, j_{t}) \notin R$. Therefore, there is no argument in $A \setminus S$ which attacks the argument $j_{t}$. If there is no argument in $S$ which attacks the argument $j_{t}$, then there is no argument in $A$ which attacks the argument $j_{t}$. It follows that the argument $j_{t}$ is defended by $S$ in $F$, a contradiction with the assumption. Thus, there is some argument $i_{r} \in S (1 \leq r \leq k)$ which attacks the argument $j_{t}$, $i.e.$, $(i_{r}, j_{t}) \in R$. This implies that $a_{i_{r}, j_{t}} = 1$, and thus the column vector $M^{^{s}}_{*, t}$ of the $s$-block $M^{i_{1}, i_{2}, ..., i_{k}}_{j_{1}, j_{2}, ..., j_{h}}$ of $S$ is non-zero.

Conversely, suppose that the conditions $(1)$ and $(2)$ are hold. Let $j_{t} \in A \setminus S (1 \leq t \leq h)$, we consider the column vector $M^{c}_{*, t}$ of the $c$-block $M^{j_{1}, j_{2}, ..., j_{h}}_{j_{1}, j_{2}, ..., j_{h}}$ of $S$. If the column vector $M^{c}_{*, t}$ is zero, then by condition $(2)$ we have that the column vector $M^{s}_{*, t}$ of the $s$-block $M^{i_{1}, i_{2}, ..., i_{k}}_{j_{1}, j_{2}, ..., j_{h}}$ of $S$ is non-zero. It follows that there is some $i_{q} \in S (1 \leq q \leq k)$ such that $a_{i_{q}, j_{t}} = 1$, $i.e.$, $(i_{q}, j_{t}) \in R$. This means that the argument $i_{q}$ attacks the argument $j_{t}$ in $F$. Considering that $S$ is a conflict-free set, there is no argument $i_{r} \in S (1 \leq r \leq k)$ which attacks the argument $i_{q}$ in $F$.

If the column vector $M^{c}_{*, t}$ is non-zero, then the row vector $M^{c}_{t, *}$ is also non-zero. By condition $(1)$, the column vector $M^{s}_{*, t}$ of the $s$-block $M^{i_{1}, i_{2}, ..., i_{k}}_{j_{1}, j_{2}, ..., j_{h}} = M^{s}$ of $S$ is zero. It follows that $a_{i_{r}, j_{t}} = 0$, $i.e.$, $(i_{r}, j_{t}) \notin R$ for each $1 \leq r \leq k$. This implies that there is no argument $i_{r} \in S (1 \leq r \leq k)$ which attacks the argument $j_{t}$ in $F$.

To sum up, we conclude that the argument $j_{t} \in A \setminus S (1 \leq t \leq h)$ is not defended by $S$. \\

\hspace{-10mm} {\bf \Large 8. Conclusions and perspectives}

\vspace{5mm} In this paper, we introduced the matrix $M(F)$ of an argumentation framework $F = (A, R)$, and the $cf$-block $M^{cf}$, $s$-block $M^{s}$, $a$-block $M^{a}$ and $c$-block $M^{c}$ of a set $S \subset A$, presented several theorems to decide various extensions (stable, admissible, complete) of the AF, by blocks of the matrix  $M(F)$  of $F$ and relations between these blocks.

Interestingly, the $s$-block $M^{s}$ ($a$-block $M^{a}$, $c$-block $M^{c}$) of $S$ corresponds to the determination for $S$ to be a stable extension
(admissible extension, complete extension respectively). And, the $c$-block of $S$ is exactly the complementary block of the $cf$-block of $S$, the $a$-block of $S$ is exactly the complementary block of the $s$-block of $S$. Furthermore, we can decide basic extensions of an argumentation framework by the special feature of blocks and relations between these blocks. These facts indicate that there is indeed a corresponding relation between the argumentation framework and its matrix. So, we can investigate the structure and properties of an argumentation framework by using the theory and method of matrix.

For the other common extension semantics (preferred, grounded, ideal, semi-stable and eager) of Dung's argumentation framework not discussed in the above sections, we can also provide the matrix method  to describe them, by combining the obtained results. For example, if we want to decide that a complete extension $S \subset A$ is grounded in $F = (A, R)$, we could first find out all the complete extensions by theorem 20. Then, we compare the $cf$-blocks of these complete extensions. If the $cf$-block of $S$ is the minimal one in the collection of $cf$-blocks of all complete extensions, then we claim that $S$ is a grounded extension.

The prospectives are that, we can find out the internal pattern of AFs and the relations between different objects which we concerned in AFs, by studying blocks of the matrix of AFs. Our future goal is to develop the matrix method in the related areas, such as argument acceptability, dialogue games, algorithmic and complexity and so on [7, 11, 8, 13, 16, 12].\\

%\end{document}

\hspace{-8mm}{\large \bf  References}
\hspace{5mm}

%\end{document}

\end{document}